\documentclass[12pt,preprint]{aastex}
\shorttitle{Light Curves of Swift Gamma Ray Bursts}
\shortauthors{Cea}

\begin{document}

   \title{LIGHT CURVES OF SWIFT GAMMA RAY BURSTS}
   \author{Paolo Cea}

   \affil{Dipartimento Interateneo di Fisica, Universit\`a di Bari,
              Via G. Amendola 173, I-70126 Bari\\
         \and
             INFN - Sezione di Bari, Via G. Amendola 173, I-70126 Bari \\
                \email{Paolo.Cea@ba.infn.it}
             }

\begin{abstract}
Recent observations from the Swift gamma-ray burst mission  indicate that a  fraction of  gamma ray  bursts are
characterized by a canonical behaviour of the X-ray afterglows. We present an effective theory which allows us to account
for X-ray light curves of both (short - long) gamma ray bursts and X-ray rich flashes. We propose that  gamma ray bursts originate from
massive magnetic powered pulsars.
\end{abstract}
   \keywords{ gamma-rays: bursts}
\section{INTRODUCTION}
\label{Introduction}
The unique capability of the Swift satellite has yielded the discovery of interesting new properties of short and long gamma ray burst
(GRB) X-ray afterglows. Indeed, recent observations have provided new informations on the early behavior ( $t \, < \, 10^3 - 10^4 \, sec$) of the X-ray
light curves of  gamma ray bursts. These early time afterglow observations revealed
that~\cite{Chincarini:2005,Nousek:2005,O'Brien:2006} a  fraction of bursts have a generic shape consisting of three
distinct segments: an initial very steep decline with time, a subsequent very shallow decay, and a final steepening (for a
recent review, see Piran 2005, Meszaros 2006). This canonical behaviour of the X-ray afterglows of gamma ray bursts is
challenging the standard relativistic fireball model, leading to several alternative models (for a recent review
of some of the current theoretical interpretations, see M\'esz\'aros 2006 and references therein). \\
\indent
In order to determine the nature of both short and long gamma ray bursts, more detailed theoretical modelling is needed to establish a clearer picture of the mechanism. In particular, it is important to have at disposal an unified, quantitative description of the
X-ray afterglow light curves.  \\
\indent
The main purpose of this paper is to present an effective theory which allows us to account for X-ray light curves of both
gamma ray bursts and X-ray rich flashes (XRF). In a recent paper~\cite{cea:2006} we set up a quite general approach to
cope with light curves from anomalous $X$-ray pulsars (AXP) and soft gamma-ray repeaters (SGR). Indeed, we find that the
canonical light curve of the X-ray afterglows is very similar to the light curve after the June 18, 2002 giant burst from
AXP 1E 2259+586 (Woods et al. 2004). This suggests that our approach can be extended also to gamma ray bursts. \\
\indent
The plan of the paper is as follows. In Sect.~\ref{light} we briefly review the  general formalism presented in Cea (2006)
to cope with light curves. After that, in Sect.~\ref{050315} through \ref{050416A} we carefully compare our theory with
the several gamma ray burst light curves. In Sect.~\ref{origin} we propose that gamma ray bursts originate from massive
magnetic powered pulsars, namely pulsars with super strong dipolar magnetic field and mass $M \sim 10 \, M_{\bigodot}$.
Finally, we draw our conclusions in Sect.~\ref{conclusion}.
\section{\normalsize{LIGHT CURVES}}
\label{light}
Gamma ray bursts may be characterized by some mechanism which dissipates injected energy in a compact region. As a
consequence the observed luminosity is time-dependent. In this section, following Cea (2006), we briefly discuss an
effective description that allows us to determine the light curves, i.e. the time dependence of the luminosity.
After that, we shall compare our approach with several light curves of Swift gamma ray bursts. \\
\indent
In general, irrespective of the details of the dissipation process, the dissipated energy  leads to the luminosity $L(t) =
- \frac{d E(t)}{dt}$. Actually, the precise behavior of $L(t)$ is determined once the dissipation mechanisms are known.
However, we may accurately reproduce the time variation of $L(t)$ without precise knowledge of the microscopic dissipative
mechanisms. Indeed, on general grounds we expect that the dissipated energy is given by:
\begin{equation}
\label{2.1}
L(t) \; \; = \; \; - \; \frac{d E(t)}{dt} \; \; = \; \kappa(t) \; E^\eta \; \; , \; \; \eta \; \leq \; 1 \; \; \; ,
\end{equation}
where $\eta$ is the efficiency coefficient. For an ideal system, where the initial injected energy is huge, the linear
regime where $\eta = 1$ is appropriate. Moreover, we may safely assume that $\kappa(t) \simeq \kappa_0$ constant. Thus we
get:
\begin{equation}
\label{2.2}
L(t) \; \; = \; \; - \; \frac{d E(t)}{dt} \; \; \simeq \; \kappa_0 \; E \; \; .
\end{equation}
It is then straightforward to solve Eq.~(\ref{2.2}):
\begin{eqnarray}
\label{2.3}
  E(t) \; & = & \; E_0 \; \exp(- \frac{t}{\tau_0}) \; \; \; \; , \; \; \;
 L(t) \;  = \;  L_0 \; \exp(- \frac{t}{\tau_0}) \; \; ,
 \\
 \nonumber
 L_0 \; & = &\;  \frac{E_0}{\tau_0} \; \; , \; \;
 \tau_0 \; = \; \frac{1}{\kappa_0} \; \; \; .
 \end{eqnarray}
Note that the dissipation time $\tau_0  = \frac{1}{\kappa_0}$ encodes all the physical information on the microscopic
dissipative phenomena. Since the injected energy is finite, the dissipation of energy degrades with the decrease in the
available energy. Thus, the relevant equation is Eq.~(\ref{2.1}) with $\eta < 1$. In this case, by solving Eq.~(\ref{2.1})
we find:
\begin{equation}
\label{2.4}
 L(t) \;  = \;  L_0 \; \left ( 1 - \frac{t}{t_{dis}} \right )^{\frac{\eta}{1-\eta}}  \; \; ,
\end{equation}
where we have introduced the dissipation time:
\begin{equation}
\label{2.5}
t_{dis}  \; \; = \; \; \frac{1}{\kappa_0} \; \frac{E_0^{1-\eta}}{1-\eta}  \; \; .
\end{equation}
Then, we see that the time evolution of the luminosity is linear up to some time $t_{break}$, and after that we have a
break from the linear regime $\eta = 1$ to a non linear regime with $\eta < 1$. If we indicate the total dissipation time
by $t_{dis}$ ,  we get:
\begin{eqnarray}
\label{2.6}
L(t) \; & = & \;  L_0 \; \exp(- \frac{t}{\tau_0}) \; \; \; \;  \; \; ,  \; \; \; \; \; \; \; \; \; \; \;
                    \; \; \; \; \; \;\; \; 0 \; < \; t \; < \; t_{break}  \; \;  ,  \\
\nonumber
L(t) \; & = & \;  L(t_{break}) \;  \left ( 1 - \frac{t-t_{break}}{t_{dis}-t_{break}} \right )^{\frac{\eta}{1-\eta}}  ,
               \;  t_{break}  \; < \; t \; < \; t_{dis}   \; \;  .
\end{eqnarray}
Equation~(\ref{2.6}) is relevant for light curves where there is a huge amount of energy to be dissipated. \\
\indent
Several observations indicate that after a giant burst there are smaller and more recurrent bursts. According to our
approach, we may think about these small bursts as similar to the seismic activity following a giant earthquake
 (for statistical similarities between bursts and earthquakes, see Cheng et al. 1995). 
 These seismic bursts are characterized by very different light curves from the giant burst light curves. \\
During these seismic bursts there is an almost continuous injection of energy, which tends to sustain an almost constant
luminosity. This corresponds to an effective $\kappa$ in Eq.~(\ref{2.1}) which decreases smoothly with time. The simplest
choice is:
\begin{equation}
\label{2.7}
\kappa(t) \; = \; \frac{\kappa_0}{1 + \kappa_1 t}  \; \; \; .
\end{equation}
Inserting this into  Eq.~(\ref{2.1}) and integrating, we get:
\begin{equation}
\label{2.8}
 E(t)  \; = \; \left [ E_0^{1-\eta} \; - (1 - \eta) \;  \frac{\kappa_0}{\kappa_1} \; \ln (1 + \kappa_1 t)
  \right ]^{\frac{\eta}{1-\eta}} \; \; \; ,
\end{equation}
so that the luminosity is:
\begin{equation}
\label{2.9}
 L(t)  \; = \; \frac{L_0}{(1 + \kappa_1 t)^{\eta}} \; \left [1  \; -  (1 - \eta) \;  \frac{\kappa_0}{\kappa_1 E_0^{1-\eta}}
          \; \ln (1 + \kappa_1 t) \right ]^{\frac{\eta}{1-\eta}} \; \; \; .
\end{equation}
After defining the dissipation time as
\begin{equation}
\label{2.10}
 \ln (1 + \kappa_1 \tau_{dis}) \; =  \;  \frac{\kappa_1}{\kappa_0 }\; \frac{E_0^{1-\eta}}{1 - \eta} \; \; \; ,
\end{equation}
we rewrite Eq.~(\ref{2.9}) as
\begin{equation}
\label{2.11}
 L(t)  \; = \; \frac{L_0}{(1 + \kappa_1 t)^{\eta}} \; \left [ 1  \; -   \;  \frac{\ln (1 + \kappa_1 t)}{\ln (1 + \kappa_1
          \tau_{dis})}  \right ]^{\frac{\eta}{1-\eta}} \; \; \; .
\end{equation}
Note that the light curve in Eq.~(\ref{2.11}) depends on two characteristic time constants $\frac{1}{\kappa_1}$ and
$\tau_{dis}$. We see that $\kappa_1 \,  \tau_{dis}$, which is roughly the number of small bursts that occurred in the
given event, gives an estimation of the seismic burst intensity. Moreover, since during the seismic bursts the injected
energy is much smaller than in giant bursts, we expect values of $\eta$ which are lower with respect to typical values in
giant bursts.
\\
\indent
As we alluded in the Introduction, the canonical light curves of the X-ray afterglows are very similar to the light curve
after the  2002 June 18 giant burst from AXP 1E 2259+586. In Cea (2006) we were able to accurately reproduce the puzzling
light curve of the June 2002 burst  by assuming that AXP 1E 2259+586 has undergone a giant burst, and soon after has
entered into intense seismic burst activity. Accordingly,  we may parameterize the X-ray afterglow light curves of gamma
ray bursts as:
\begin{equation}
\label{2.12}
 L_{GRB}(t) \; = \;  L_{G}(t) \;  + \; L_{S}(t) \; \; \; \; ,
\end{equation}
where, since there are no available data during the first stage of the outbursts, we have for the giant burst's
contribution:
\begin{equation}
\label{2.13}
L_{G}(t) \; = \;  L_{G}(0) \;  \left ( 1 - \frac{t}{t_{diss}} \right )^{\frac{\eta_{G}}{1-\eta_{G}}} \; \; ,
               \; \; \;  0 \; < \; t \; < \; t_{diss}  \; \;  ,
\end{equation}
while $L_{S}(t)$ is given by:
\begin{equation}
\label{2.14}
 L_{S}(t)  = \frac{L_{S}(0)}{(1 + \kappa t)^{\eta_{S}}} \;
          \left [ 1  \; -   \;  \frac{\ln (1 + \kappa t)}{\ln (1 + \kappa
           \tau_{diss})}  \right ]^{\frac{\eta_{S}}{1-\eta_{S}}}   , \; \;
                 0 \; < \; t \; < \; \tau_{diss} \; \; .
\end{equation}
Note that, unlike the anomalous $X$-ray pulsars  and soft gamma-ray repeaters, we do not need to take care of the
quiescent flux since the gamma ray burst sources are at cosmological distances. \\
\indent
In the  following Sections we select  a collection of GRBs with the aim to  illustrate the variety of displayed light curves.  In general, we reproduce the data of light curves  from the  original figures. For this reason, we display the light curves with the same time intervals as in the original figures. So that, lacking the precise values  of data the best fits to our light curves are  only indicative. In view of this, a quantitative comparison  with different models is not possible. The unique exception is  GRB 050801 were the data was taken from Table~1 in  Rykoff et al. (2006).  In that case (see Sect.~\ref{050801})  we  indicate the reduced chisquare.
\subsection{\normalsize{GRB 050315}}
\label{050315}
On  2005  March 15 the Swift Burst Alert Telescope (BAT) triggered  and located on-board GRB 050315~\cite{Vaughan:2006}.
After about $80 \, sec$ the Swift X-ray Telescope (XRT) began observations until about $10 \, days$, providing one of the
best-sampled X-ray light curves of a gamma ray burst afterglow. \\

\vspace{0.6 cm}
\begin{figure}[ht]
   \centering
   \epsscale{0.7}
  \plotone{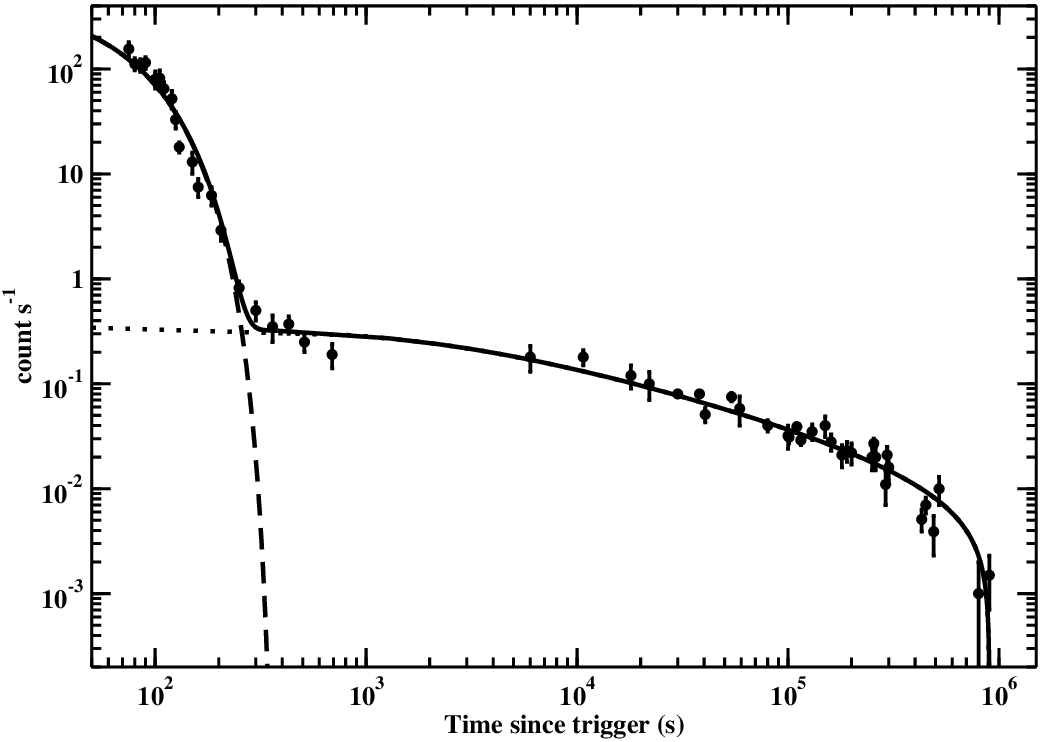}
\caption{\label{fig_1}
Light curve of GRB 050315 in the $0.2 - 5 \, keV$ band. The data was extracted from Fig.~5 in Vaughan et al. (2006). The
full line is our light curve Eq.~(\ref{2.12}); the dashed and dotted lines are  Eq.~(\ref{2.13}) and Eq.~(\ref{2.14})
respectively, with parameters in Eq.~(\ref{2.15}).}
\end{figure}
\indent
In Fig.\ref{fig_1} we display the light curve of GRB 050315 in the $0.2 - 5 \, keV$ band. The data was extracted from
Fig.~5 in Vaughan et al.(2006). \\
\indent
A tentative fit to the X-ray light curve within the standard  relativistic fireball model has been proposed in Granot et
al. (2006) using a two-component jet model. An alterative description of the light curve of GRB 050315 within the
cannonbal model is presented in Dado \& De Rujula (2005). \\
We fitted the data to our light curve Eq.~(\ref{2.12}). Indeed, we find a rather good description of the data with the
following parameters (see Fig.~\ref{fig_1}):
\begin{eqnarray}
\label{2.15}
 L_{G}(0) \; & \simeq & \; 5.2 \;  10^{2} \; \frac{count}{ sec} \; \; , \; \; \; \;
  \; \; \;
\eta_{G} \;  \simeq  \;
             0.867 \; \; , \; \; t_{diss} \;  \simeq \;  380 \; sec \;  \; \; \; \; \; \; \; \; \\
\nonumber
 L_{S}(0) \; & \simeq & \; 0.35 \; \frac{count}{sec} \; \; \;  , \; \; \; \; \; \; \;
  \; \;
  \eta_{S} \;  \simeq  \; 0.4
  \; ,  \; \tau_{diss} \;  \simeq \; 9.0 \; 10^5 \; sec \;  ,  \; \kappa  \; \simeq \; 5.0 \; 10^4 \; sec^{-1}  \; .
\end{eqnarray}
A few comments are in order.  As discussed in the Sect.~\ref{light}, since  we lack the precise values  of data, a quantitative comparison  
 of our light curve with data is not   possible. Nevertheless,   Fig.~\ref{fig_1} shows that the agreement with data is rather good. Moreover, our efficiency exponents $\eta_{G}$ and $ \eta_{S}$ are consistent with the values found in giant bursts from  anomalous $X$-ray pulsars and soft gamma-ray repeaters~\cite{cea:2006}. Note that, as expected, we have $ \eta_{S} < \eta_{G}$ .
\subsection{\normalsize{GRB 050319}}
\label{050319}
Swift discovered GRB 050319 with the Burst Alert Telescope and began observing after $225 \, s$ after the burst
onset~\cite{Cusumano:2006}.  \\
\begin{figure}[ht]
   \centering
   \epsscale{0.7}
 \plotone{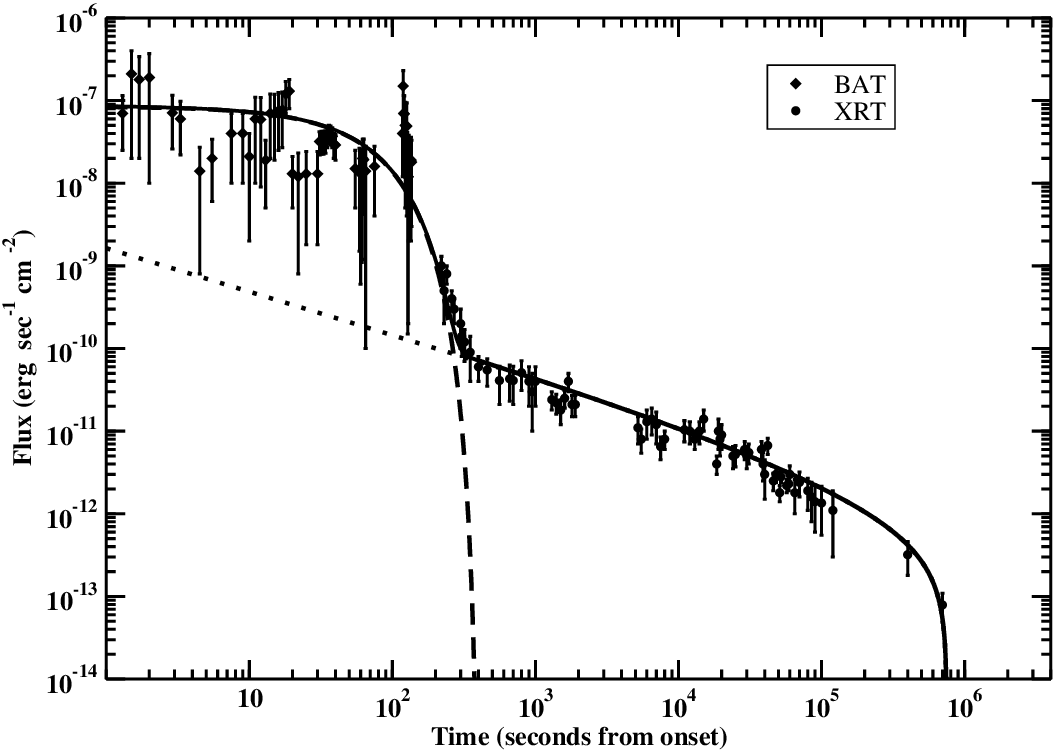}
\caption{\label{fig_2}
XRT light curve of GRB 050319 in the $0.2 - 10 \, keV$ band. The data was extracted from Fig.~2 in Cusumano et al. (2006).
The BAT light curve was obtained by extrapolating the BAT count rate into the XRT energy band with the best-fit spectral
model~\cite{Cusumano:2006}. The full line is our light curve Eq.~(\ref{2.12}); the dashed and dotted lines are
Eq.~(\ref{2.13}) and Eq.~(\ref{2.14}) respectively, with parameters in Eq.~(\ref{2.16}).}
\end{figure}
\indent
The X-ray afterglow was monitored by the XRT up to 28 days after the burst. In Fig.~\ref{fig_2} we display the X-ray light
curve in the $0.2 - 10 \, keV$ band. The data are extracted from Fig.~2 in Cusumano et al. (2006). Note that the light
curve in the early stage of the outflow has been obtained  extrapolating the BAT light curve in the XRT band by using the
the best-fit spectral model~\cite{Cusumano:2006}. \\
\indent
An adeguate description of the XRT light curve of GRB 050319 within the cannonbal model is presented in Dado \& De Rujula
(2005). However, we note that the extrapolation of the best-fit light curve towards the first stage of the outburst
overestimates the observed flux by orders of magnitude. On the other hand, we may easily account for the observed flux
decay by our light curve. Indeed, in Fig.~\ref{fig_2} we compare our light curve Eq.~(\ref{2.12}) with observational data.
The agreement is quite satisfying, even during the early-time of the outburst, if we take:
\begin{eqnarray}
\label{2.16}
 L_{G}(0) \; & \simeq & \; 8.5 \;  10^{-8} \; \frac{erg }{ cm^{2} \, sec} \; \; , \; \; \; \;
  \;
 \eta_{G} \; \simeq \;
             0.867 \; \; , \; \; t_{diss} \;  \simeq \;  410 \; sec \;  \; \; \; \; \\
\nonumber
 L_{S}(0) \; & \simeq & \; 5.5 \;  10^{-9} \; \frac{erg}{ cm^{2} \, sec}\; \; \; ,   \; \; \;
  \; \;
  \eta_{S} \; \simeq \; 0.45
  \; ,  \; \tau_{diss} \;  \simeq \; 7.5 \; 10^5 \; sec \;  ,  \; \kappa  \; \simeq \; 10 \; sec^{-1}  \; .
\end{eqnarray}
\subsection{\normalsize{XRF 050406}}
\label{050406}
On  2005 April 6 BAT triggered on GRB 050406~\cite{Romano:2006a}. The gamma-ray characteristics of this burst, namely the
softness of the observed spectrum and the absence of significant emission above $\sim \, 50 \, keV$, classify the burst as
an X-ray flash (XRF 050406). \\
\indent
\begin{figure}[ht]
   \centering
   \epsscale{0.7}
   \plotone{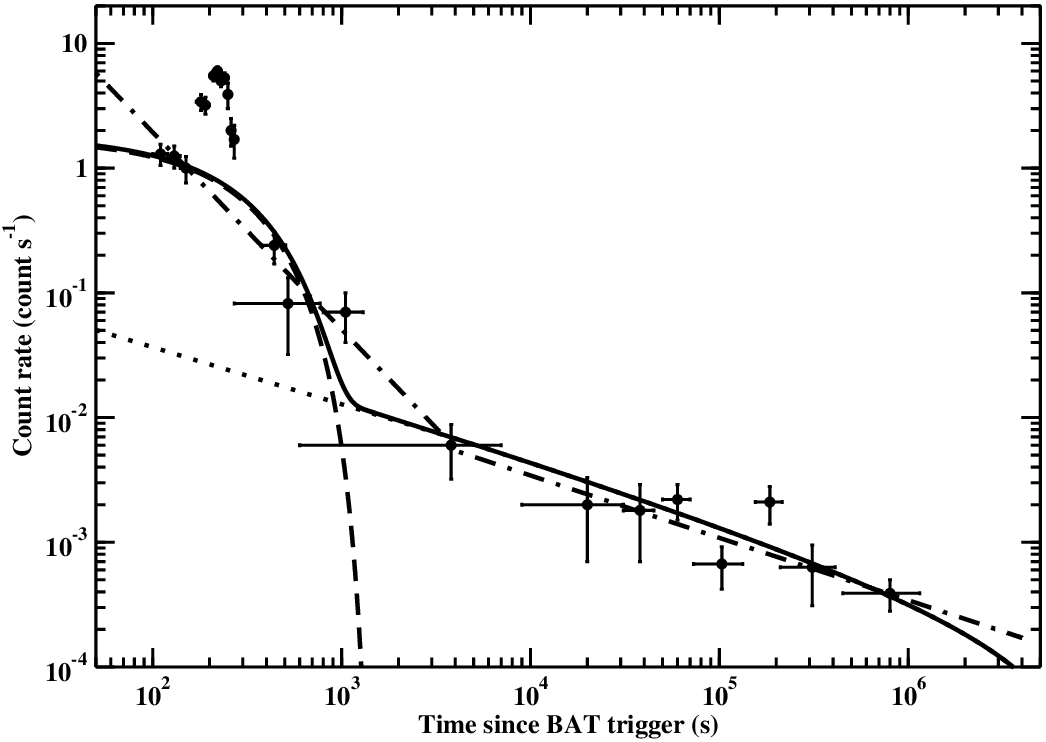}
\caption{\label{fig_3}
X-ray light curve of XRF 050406 in the $0.2 - 10 \, keV$ energy band. The data was extracted from Fig.~2 in Romano et al.
(2006a). The dot-dashed line is the broken power-law best fit~\cite{Romano:2006a}. The full line is our light curve
Eq.~(\ref{2.12}); the dashed and dotted lines are Eq.~(\ref{2.13}) and Eq.~(\ref{2.14}) respectively, with parameters in
Eq.~(\ref{2.17}).}
\end{figure}
In Fig.~\ref{fig_3} we display the time decay of the flux. The data was taken from Fig.~2 in Romano et al. (2006a). We fit
our light curve Eq.~(\ref{2.12}) to the available data. Indeed, we find that our light curve, with parameters given by:
\begin{eqnarray}
\label{2.17}
 L_{G}(0) \;  & \simeq & \; 1.75 \; \; \frac{count}{sec} \; \; ,  \; \; \;
 \eta_{G} \; \simeq \;
             0.839 \; \; , \; \; t_{diss} \;  \simeq \;  1.5 \; 10^3 \; sec \;  \; \; \; \; \; \\
\nonumber
 L_{S}(0) \; & \simeq & \; 0.15  \; \frac{count}{sec}\; \; \; ,  \; \; \;
  \eta_{S} \; \simeq \; 0.40
  \; ,  \; \tau_{diss} \;  \simeq \; 8.0 \; 10^6 \; sec \;  ,  \; \kappa  \; \simeq \; 0.2 \; sec^{-1}  \; ,
\end{eqnarray}
allows quite a satisfying description of the decline of the flux (see Fig.~\ref{fig_3}). Note that in the fit we exclude the bump at $ t  \sim  
 200 \, sec$. For completeness, we also display in Fig.~\ref{fig_3} the phenomenological best-fit broken power 
 law~\cite{Romano:2006a}. It is worthwhile to observe that
the bump in the flux at $t \, \sim \, 200 \, sec$ is similar to the April 18, 2001 flare from SGR
1900+14~\cite{Feroci:2003}. Indeed, within our approach we believe that the bump in the flux could naturally be explained
as fluctuations in the intense burst activity (see Sect.~5.2 in Cea 2006).
\subsection{\normalsize{GRB 050801}}
\label{050801}

\begin{figure}[ht]
   \centering
   \epsscale{0.7}
   \plotone{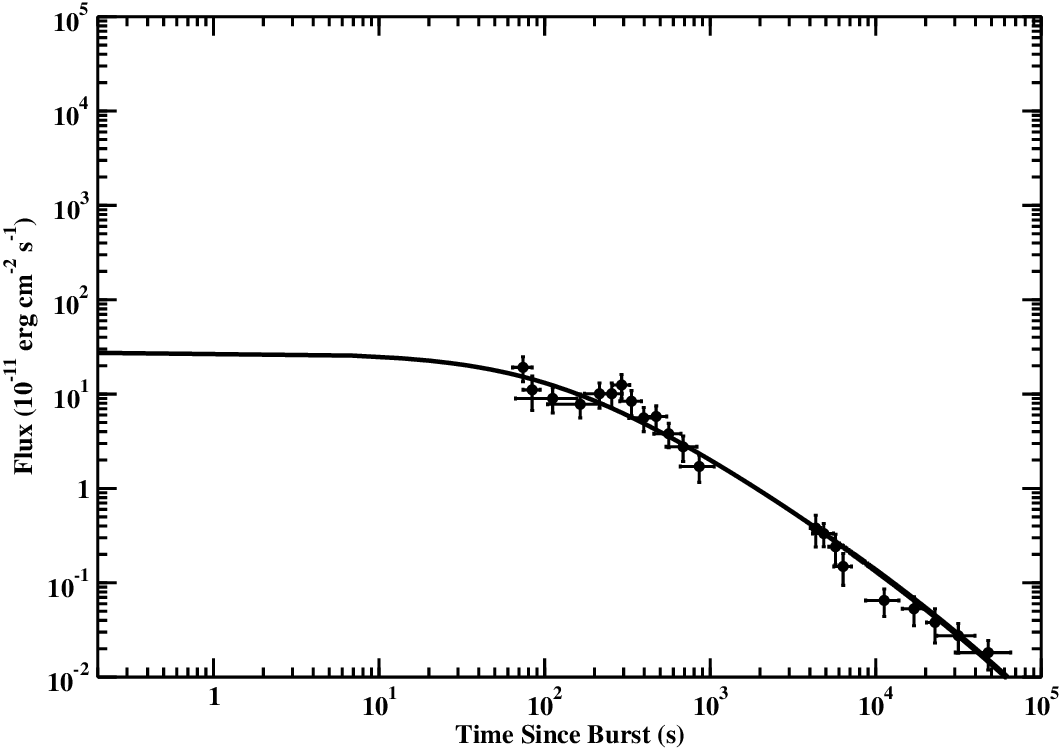}
\caption{\label{fig_4}
X-ray light curve of GRB 050801. The data was taken from Table~1 in Rykoff et al. (2006). The full line is our best-fit
light curve Eq.~(\ref{2.14})  with parameters in Eq.~(\ref{2.20}).
 }
\end{figure}

The Swift XRT obtained observations starting at 69 seconds after the burst onset of GRB 050801~\cite{Rykoff:2006}. In
Fig.~\ref{fig_4} we display the flux decay, where the data has been extracted from Table~1 in Rykoff et al. (2006). \\
\indent
In this case we are able to best fit our light curve Eq.~(\ref{2.12}) to the available data. Since the observations start
from $t \, > 74 \, s$, we perform the fit of data to the seismic burst light curve  $F_{S}(t)$, Eq.~(\ref{2.14}). To get a
sensible fit we fixed the dissipation time to $\tau_{diss} \, = \, 2.0 \, 10^6 \, s$ and $\kappa \, = \, 10^{-2} \,
s^{-1}$. The best fit of our light curve to data gives:
\begin{equation}
\label{2.20}
 L_{S}(0) \;  = \; (27.4 \; \pm \; 7.2) \; 10^{-11} \; \frac{erg }{ cm^{2} \, sec} \;  , \; \;
  \eta_{S} \; = \; 0.748 \; \pm \; 0.026 \;  \; ,
\end{equation}
with a reduced $\chi^2/dof \, \simeq \, 0.93$. In Fig.~\ref{fig_4} we compare our best-fitted light curve with
data. We see that our theory allows a satisfying description of the light curve of GRB 050801. On the other hand, it is
difficult to explain the peculiar behaviour of the light curve with standard models of early afterglow emission without
assuming that there is continuous late time injection of energy into the afterglow~\cite{Rykoff:2006}.
\subsection{\normalsize{GRB 051221A}}
\label{051221}

\begin{figure}[ht]
   \centering
   \epsscale{0.7}
   \plotone{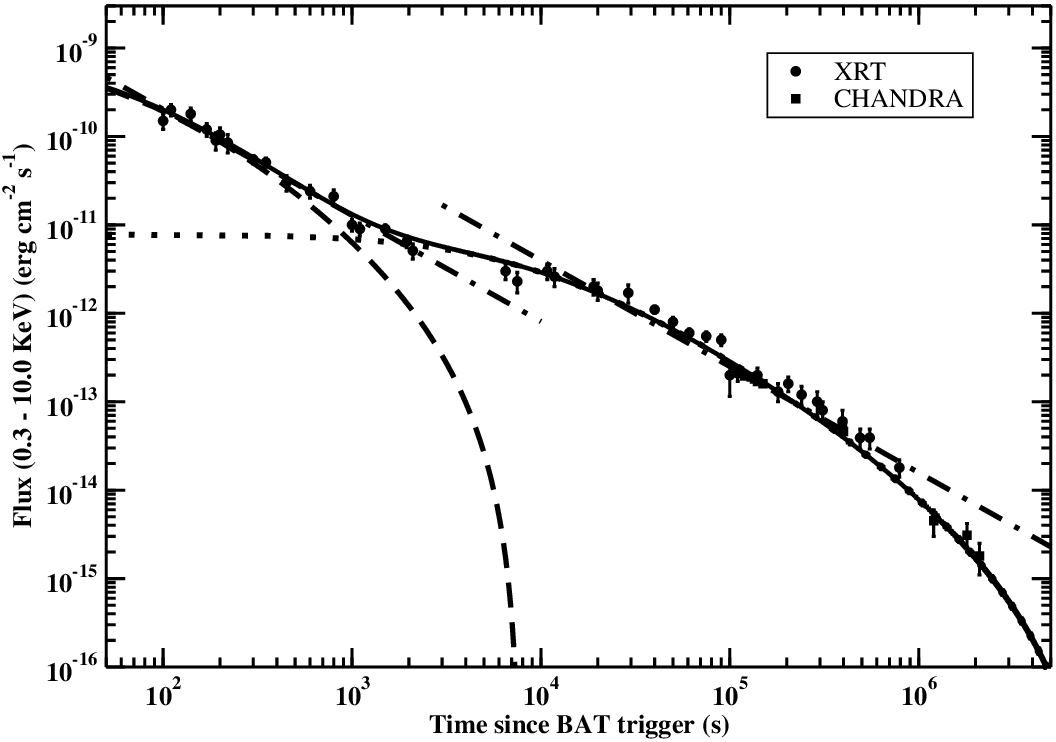}
\caption{\label{fig_5}
Combined XRT and CHANDRA light curve of the afterglow of GRB 051221A. The data was extracted from Fig.~2 in Burrows et al.
(2006). The dot-dashed lines are the phenomelogical power law fits $t^{-1.20}$. The full line is our light curve
Eq.~(\ref{2.22}), the dashed line is $L_{S_1}(t)$, and the dotted line is  $L_{S_2}(t)$.}
\end{figure}

GRB 051221A was detected by the Swift BAT on 2005 December 21. The Swift XRT observations began 88 seconds after the BAT
trigger. The late X-ray afterglow of GRB 051221A has been also observed by the Chandra ACIS-S instrument. The combined
X-ray light curve, displayed in Fig.~\ref{fig_5}, was extracted from Fig.~2 in Burrows et al. (2006).\\
\indent
From Fig.~\ref{fig_5}, we see that the combined X-ray light curve is similar to those commonly observed in long gamma ray
bursts.  However, we find that the this peculiar light curve could be interpreted within our approach as the
superimposition of two different seismic bursts. Accordingly, we may account for the  X-ray afterglow light curve of GRB
051221A by:
\begin{equation}
\label{2.22}
 L_{GRB}(t) \; = \;  L_{S_1}(t) \;  + \; L_{S_2}(t) \; \; \; \; .
\end{equation}

Indeed, we find that our light curve Eq.~(\ref{2.22}) allows a rather good description of the data once the parameters
are:
\begin{eqnarray}
\label{2.22_bis}
 L_{S_1}(0)  & \simeq &  9.8 \; 10^{-10} \; \frac{erg }{ cm^{2} \, sec} , \;
  \eta_{S_1} \simeq  0.74
  ,  \; \tau_{diss_1} \;  \simeq \; 7.8 \; 10^3 \; sec \;  ,  \; \kappa_1  \; \simeq \; 2.3 \;  10^{-2} \; sec^{-1}
   \; , \\
\nonumber
 L_{S_2}(0) \; & \simeq & \; 8.0  \; 10^{-12} \; \frac{erg }{ cm^{2} \, sec}   \; , \;
  \eta_{S_2} \; \simeq \; 0.40
  \; ,  \; \tau_{diss_2} \;  \simeq \; 1.0 \; 10^7 \; sec \;  ,  \; \kappa_2  \; \simeq \;  1.4 \;  10^{-4} \; sec^{-1}
   \; .
\end{eqnarray}
It is worth mentioning that the data displayed in Fig.~\ref{fig_5} start at $ t > 10^2 \, sec$. So that we cannot reliably 
determine the eventual giant burst contribution.  On the other hand, this peculiar light curve is well described by two different seismic bursts, much like  the intense burst activity in  anomalous $X$-ray pulsars and soft gamma-ray repeaters .
Note that   the phenomelogical power law fits overestimate the light curve for  $ t > 10^6 \, sec$.
\subsection{\normalsize{GRB 050505}}
\label{050505}

\begin{figure}[ht]
   \centering
   \epsscale{0.7}
   \plotone{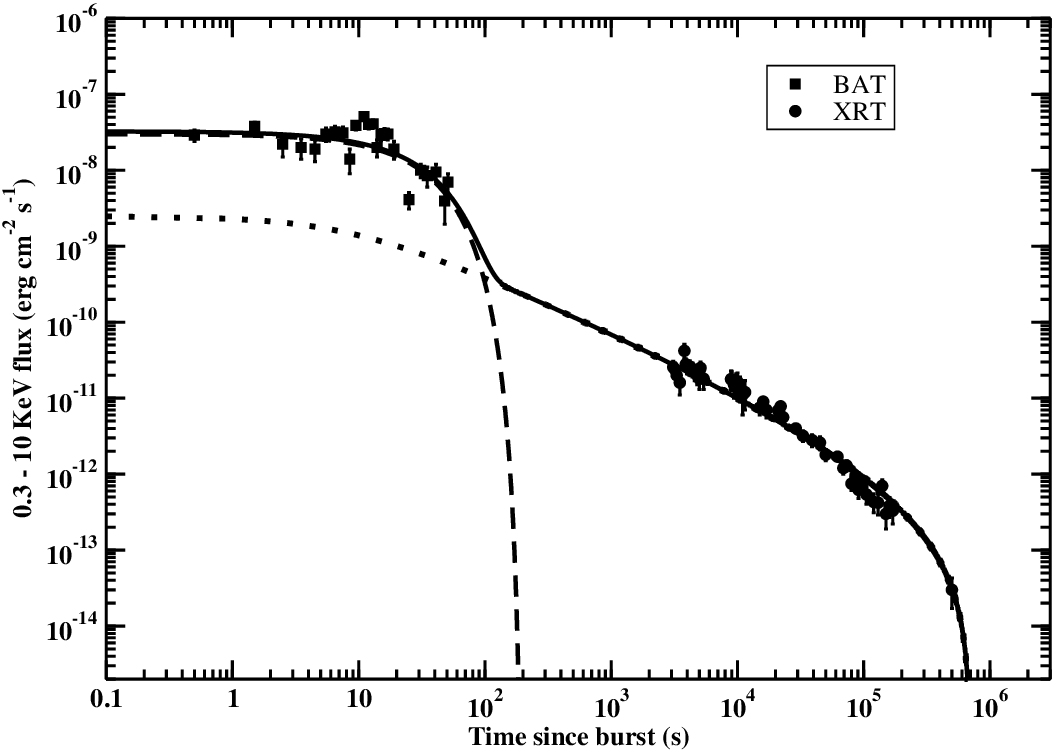}
\caption{\label{fig_6}
The combined BAT-XRT flux light curve of GRB 050505, extrapolated into the $0.3 - 10.0 \, keV$ range. The data was
extracted from Fig.~5 in Hurkett et al. (2006). The full line is our light curve Eq.~(\ref{2.12}), the dashed line is
Eq.~(\ref{2.13}), and the dotted line is Eq.~(\ref{2.14}).}
\end{figure}

On 2005 May 5 the Swift BAT triggered GRB 050505. The X-ray telescope XRT began taking data about 47 minutes after the
burst trigger. In  Fig.~\ref{fig_5} we report the combined XRT and BAT light curve of the afterglow of GRB 050505. The
data was extracted from Fig.~5 in  Hurkett et al. (2006). The BAT data were extrapolated into the the XRT band using the
best fit power law model derived from the BAT data alone~\cite{Hurkett:2006}. \\
\indent
Within the  standard models of early afterglows, the light curve is modelled by a broken power law. Nevertheless, we find
that our light curve, Eq.~(\ref{2.12}), with parameters given by:
\begin{eqnarray}
\label{2.23}
 L_{G}(0) \; & \simeq & \; 3.0 \; 10^{-8} \; \frac{erg}{ cm^{2} \, sec} \; \; , \; \; \;
 \eta_{G} \; \simeq \;
             0.867 \; \; , \; \; t_{diss} \;  \simeq \;  2.0 \; 10^2 \; sec \;  \; \; \;  \\
\nonumber
 L_{S}(0) \; & \simeq & \; 2.5 \; 10^{-9}   \; \frac{erg}{ cm^{2} \, sec} \;  , \; \;  \; \; \;
  \eta_{S} \; \simeq \; 0.58
  \; ,  \; \tau_{diss} \;  \simeq \; 7.0 \; 10^5 \; sec \;  ,  \; \kappa  \; \simeq \; 0.13 \; sec^{-1}  \; ,
\end{eqnarray}
is able to descrive quite well the X-ray afterglow (see Fig.~\ref{fig_6}).
\subsection{\normalsize{GRB 050713A}}
\label{050713}

\begin{figure}[ht]
   \centering
   \epsscale{0.7}
    \plotone{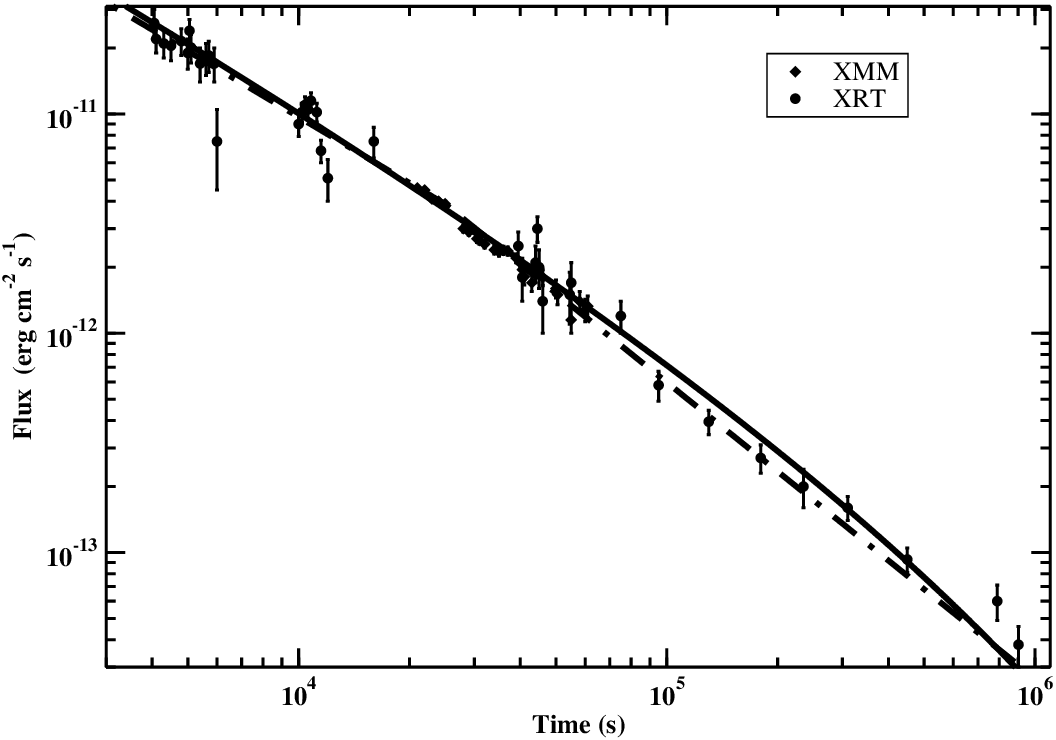}
\caption{\label{fig_7}
Joint Swift XRT and XMM-Newton light curve of the afterglow of GRB 050713A. The data was extracted from Fig.~7 in Morris
et al. (2006). The dot-dashed line is the best-fit broken power-law (\cite{Morris:2006}). The full line is our light curve
Eq.~(\ref{2.24}) with parameters in Eq.~(\ref{2.24_bis}).}
\end{figure}

In  Fig.~\ref{fig_7} we report the combined XRT and XMM-Newton light curve of the afterglow of GRB 050713A. The data was
extracted from Fig.~7 in  Morris et al. (2006). The dot-dashed line is the broken-power law best fit of the combined X-ray
light curve~\cite{Morris:2006}. \\
\indent
Within our approach we may reproduce the X-ray afterglow of GRB 050713A by our Eq.~(\ref{2.12}). However, since the giant
burst contribution to the light curve $L_G(t)$ lasts up to $t \, \sim 10^2 - 10^3 \, sec$, we need to consider only
$L_S(t)$. So that we are lead to:
\begin{equation}
\label{2.24}
 L_{GRB}(t) \; = \;  L_{S}(t) \;  \; \; .
\end{equation}
Indeed, even in this case our light curve, with parameters fixed to:
\begin{equation}
\label{2.24_bis}
 L_{S}(0) \,  \simeq \, 5.5 \, 10^{-9} \, \frac{erg }{ cm^{2} \, sec} \, , \;
 \eta_{S} \, \simeq \, 0.71  \; , \; \kappa \,  \simeq \,  7.0 \, 10^{-2} \, sec^{-1}  \; , \;
 \tau_{diss} \,  \simeq \, 9.0 \, 10^6 \, sec \;  ,
\end{equation}
reproduces quite accurately the phenomelogical broken-power law best fit (see Fig.~\ref{fig_7}).
\subsection{\normalsize{GRB 051210}}
\label{051210}

\begin{figure}[ht]
    \centering
   \epsscale{0.7}
  \plotone{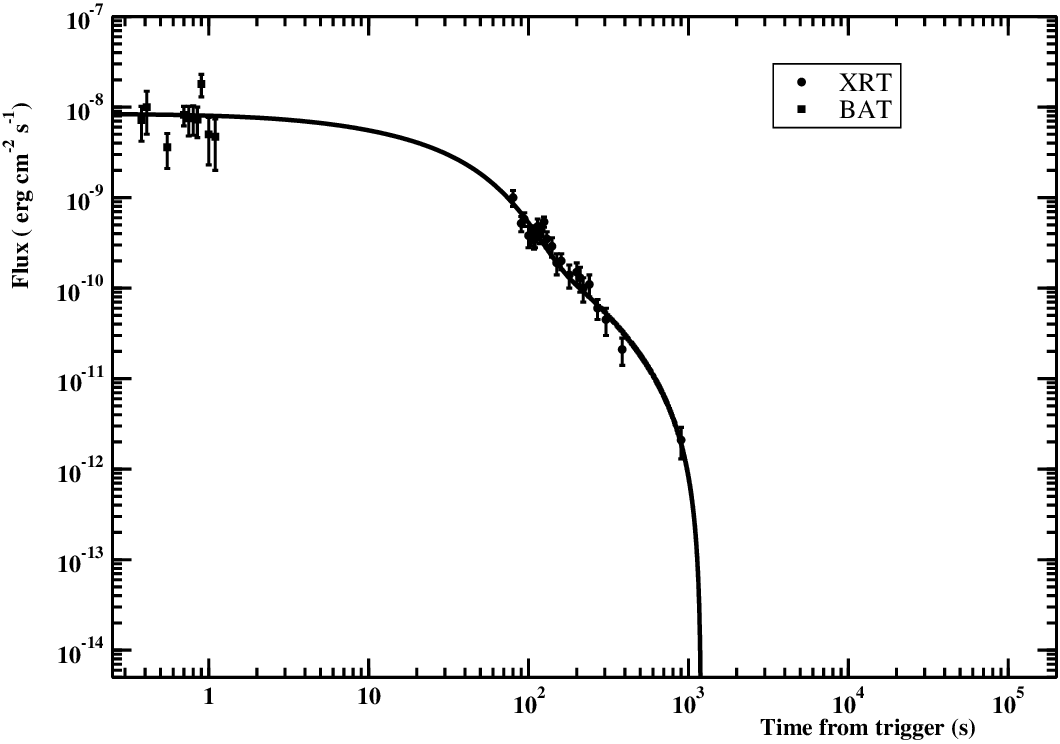}
\caption{\label{fig_8}
XRT  light curve  of GRB 051210. The data was extracted from Fig.~2 in La Parola et al. (2006). The BAT light curve was
extrapolated into the XRT energy band with the best-fit spectral model~\cite{LaParola:2006}. The full line is our light
curve Eq.~(\ref{2.12}) with parameters in Eq.~(\ref{2.25}).}
\end{figure}

GRB 051210 triggered the Swift BAT on  2005 December 12. The burst was classified as short gamma ray burst. In Fig.~2 in
La Parola et al. (2006) it is presented the XRT light curve decay of GRB 051210. The BAT light curve was extrapolated into
the $0.2 - 10 \, keV$ band by converting the BAT count rate with the factor derived from the BAT spectral parameters. \\
\indent
In  Fig.~\ref{fig_8} we report the combined BAT and XRT light curve of the afterglow of GRB 051210. The data was extracted
from Fig.~2 in  La Parola et al. (2006). In Fig.~\ref{fig_8} we also display our best fit light curve Eq.~(\ref{2.12})
with parameters:
\begin{eqnarray}
\label{2.25}
 L_{G}(0) \; & \simeq & \; 4.5 \; 10^{-9} \; \frac{erg}{ cm^{2} \, sec} \; \; , \; \; \;
 \eta_{G} \; \simeq \;
             0.867 \; \; , \; \; t_{diss} \;  \simeq \;  2.8 \; 10^2 \; sec \;  \; \; \; \; \; \; \; \\
\nonumber
 L_{S}(0) \; & \simeq & \; 4.0 \; 10^{-9}   \; \frac{erg}{ cm^{2} \, sec} \; \;  ,   \; \; \;
  \eta_{S} \; \simeq \; 0.63
  \; ,  \; \tau_{diss} \;  \simeq \; 1.2 \; 10^3 \; sec \;  ,  \; \kappa  \; \simeq \; 0.1 \; sec^{-1}  \; .
\end{eqnarray}
Even in this case the agreement between our light curve and the data is satisfying.
\subsection{\normalsize{GRB 060121}}
\label{060121}

\begin{figure}[ht]
   \centering
   \epsscale{0.7}
 \plotone{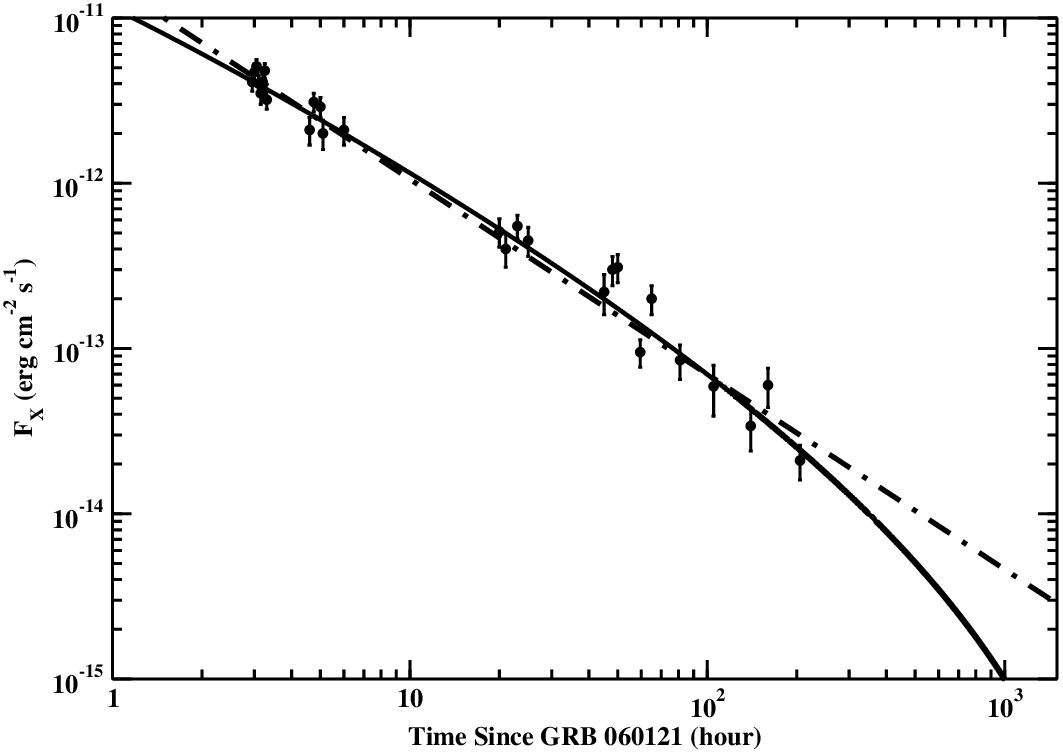}
\caption{\label{fig_9}
X-ray ($0.2 - 10 \, keV$) light curve of GRB 060121. The data was extracted from Fig.~1 in Levan et al. (2006). The
dot-dashed line is the phenomelogical power-law fit~\cite{Levan:2006}. The full line is our light curve Eq.~(\ref{2.24})
with parameters in Eq.~(\ref{2.26}).}
\end{figure}

GRB 060121 was detected by HETE-2 on January 21, 2006. Swift performed observations beginning at January 22,
2006~\cite{Levan:2006}. GRB 060121 was identified as a short and spectrally hard burst. \\
\indent
In Fig.~\ref{fig_9} we report the X-ray light curve in the $0.3 - 10.0 \, keV$ band. The data has been extracted from
Fig.~1 in Levan et al. (2006).  We also display the phenomenological power-law best fit $L(t) \, \sim \,
t^{-1.18}$~\cite{Levan:2006}. \\
\indent
Within our approach we may reproduce the X-ray afterglow of GRB 050713A by our Eq.~(\ref{2.12}). Even in this case we need
to consider only the seismic burst contribution $L_S(t)$. Indeed, we find that our Eq.~(\ref{2.24}) reproduces quite
accurately the phenomelogical power law best fit with the following parameters (see Fig.~\ref{fig_9}):
\begin{equation}
\label{2.26}
 L_{S}(0) \;  \simeq \; 9.5 \; 10^{-11} \; \frac{erg }{ cm^{2} \, sec} \; ,  \;
 \eta_{S} \; \simeq \; 0.70 \; , \; \kappa \;  \simeq \;  8.0 \; hour^{-1} \;
 \tau_{diss} \;  \simeq \; 3.0 \; 10^3 \; hour \;  .
\end{equation}
\subsection{\normalsize{GRB 060124}}
\label{060124}
Swift BAT triggered on a precursor of GRB 060124 on 2006 January 24, about 570 seconds before the main burst. So that GRB
060124 is the first event for which there is a clear detection of both the prompt and the afterglow
emission~\cite{Romano:2006b}.  \\
\begin{figure}[ht]
   \centering
   \epsscale{0.7}
  \plotone{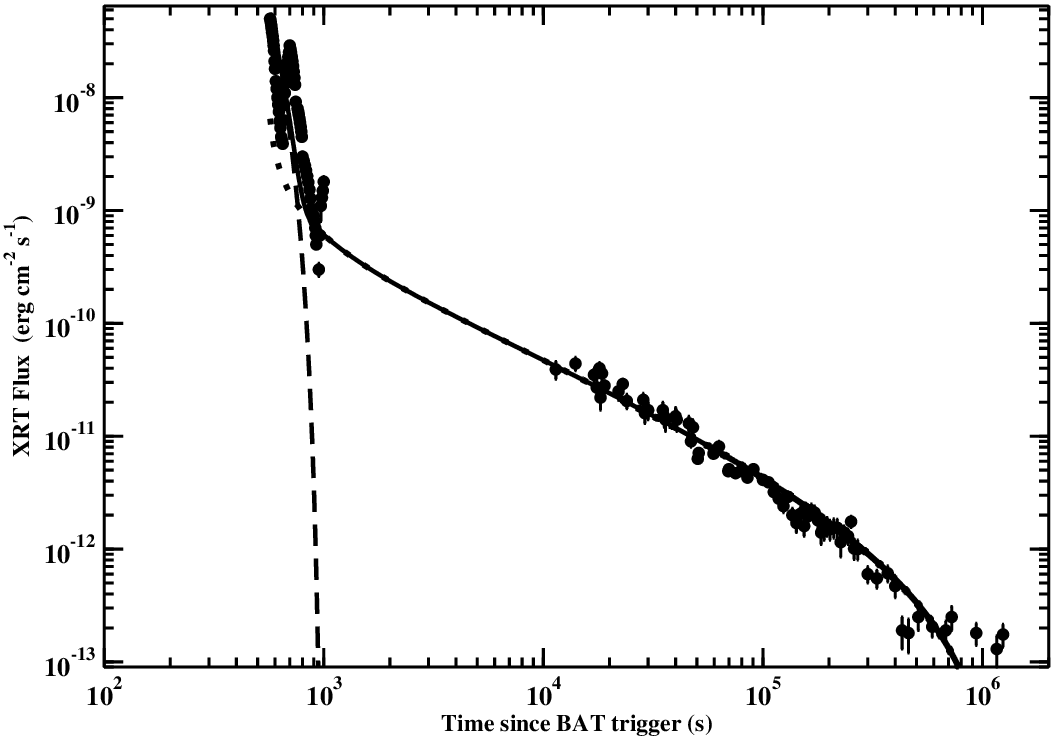}
\caption{\label{fig_10}
X-ray light curve of GRB 060124 in the $0.2 - 10 \, keV$ energy band. The data was extracted from Fig.~9 in Romano et al.
(2006b). The full line is our light curve Eq.~(\ref{2.12}); the dashed and dotted lines are Eq.~(\ref{2.13}) and
Eq.~(\ref{2.14}) respectively, with parameters given in Eq.~(\ref{2.27}).}
\end{figure}
\indent
In Fig.~\ref{fig_10} we report the X-ray light curve in the $0.3 - 10.0 \, keV$ band. The data has been extracted from
Fig.~9 in Romano et al. (2006b). In Fig.~\ref{fig_10} we display our best fit light curve Eq.~(\ref{2.12}) with
parameters:
\begin{eqnarray}
\label{2.27}
 L_{G}(0) \;  & \simeq & \; 5.0 \; 10^{-8} \; \frac{erg}{ cm^{2} \, sec} \; , \;
 \eta_{G} \; \simeq \;
             0.867 \; \; , \; \; t_{diss} \, + \, t_0 \;  \simeq \;  1.0 \; 10^3 \; sec \;  \; \; \; \; \\
\nonumber
 L_{S}(0) \; & \simeq & \; 6.5 \; 10^{-9}   \; \frac{erg}{ cm^{2} \, sec} \;  ,   \;
  \eta_{S} \; \simeq \; 0.60 \; ,  \; \tau_{diss}  \, + \, t_0  \;  \simeq \; 1.2 \; 10^6 \; sec \;  ,
  \; \kappa  \simeq 5.0 \; 10^{-2} \; sec^{-1} ,
\end{eqnarray}
where we assumed that the burst started at $t_0 \, = \, 570 \, s$. Note that our light curve interpolates the X-ray peaks
at the early stage of the outflow. On the other hand, our light curve mimics quite well the broken power-law best fit to
the XRT data (compare our Fig.~\ref{fig_10} with  Romano et al. (2006b), Fig.~9).
\subsection{\normalsize{GRB 060218}}
\label{060218}

\begin{figure}[ht]
   \centering
   \epsscale{0.7}
   \plotone{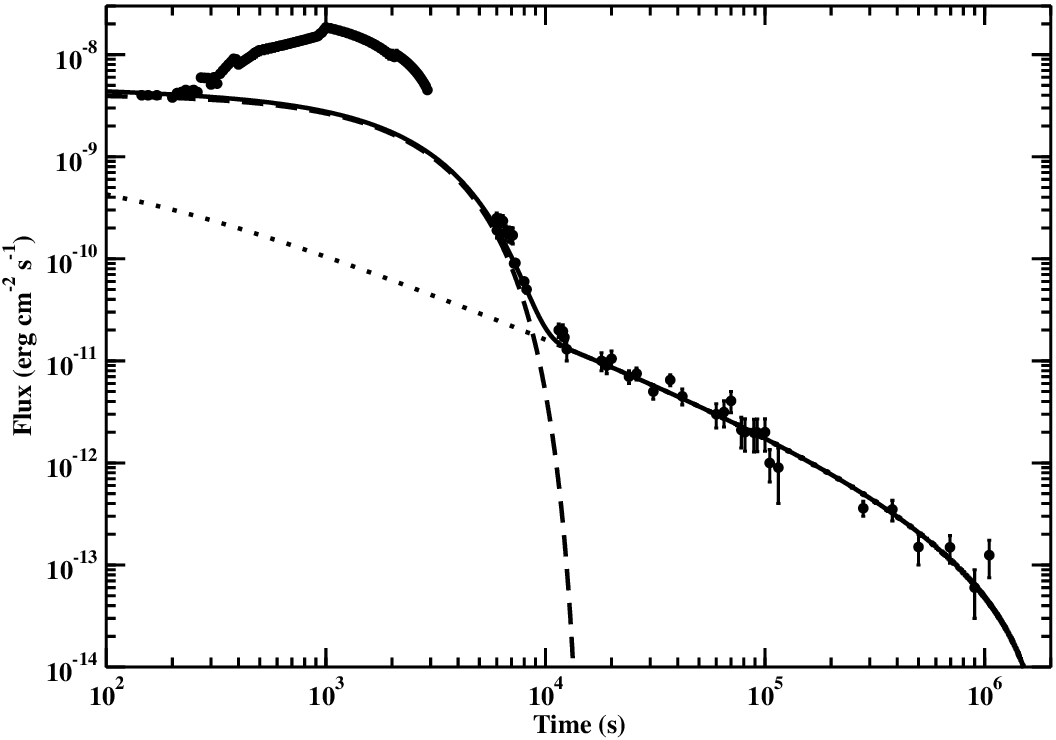}
\caption{\label{fig_11}
XRT light curve ($0.2 - 10 \, keV$) of GRB 060218. The data was extracted from Fig.~2 in Campana et al. (2006). The full
line is our light curve Eq.~(\ref{2.12}); the dashed and dotted lines are Eq.~(\ref{2.13}) and Eq.~(\ref{2.14}),
respectively.}
\end{figure}
GRB 060218 was detected with the BAT instrument on 2006 February 18. XRT began observations  159 seconds after the burst
trigger~\cite{Campana:2006}. \\
\indent
The XRT light curve is shown in Fig.~\ref{fig_11}. The data was extracted from Fig.~2 in Campana et al. (2006). We try to
interpret the XRT light curve with our light curve Eq.~(\ref{2.12}). The result of our best fit is displayed in
Fig.~\ref{fig_11}. Excluding  the data of the bump  from $ t \sim 200 \, sec$ to $ t \sim 3000 \, sec$,   the parameters for our best fit 
light curve are:
\begin{eqnarray}
\label{2.28}
 L_{G}(0) \; & \simeq & \; 4.1 \; 10^{-9} \; \frac{erg}{ cm^{2} \, sec} \; \; , \; \; \;
 \eta_{G} \; \simeq \;
             0.867 \; \; , \; \; t_{diss} \;  \simeq \;  1.55 \; 10^4 \; sec \;  \; \; \; \\
\nonumber
 L_{S}(0) \; & \simeq & \; 8.0 \; 10^{-10}   \; \frac{erg}{ cm^{2} \, sec} \; ,   \; \;
  \eta_{S} \; \simeq \; 0.60 \; ,  \; \tau_{diss} \;  \simeq \; 2.0 \; 10^6 \; sec \;  ,
  \; \kappa  \; \simeq \; 1.3 \; 10^{-2} \; sec^{-1}  \; ,
\end{eqnarray}
Indeed, our light curve is able to reproduce quite well the data. However, there is a clear excess in the observed light
curve with respect to our light curve in the early-time afterglow. We believe that this excess is due to a component which
is not directly related to the burst. Indeed, Campana et al. (2006) pointed out that there was a soft component in the
X-ray spectrum, that is present in the XRT starting from 159 s up to about $10^4$ s. This soft component could be
accounted for by a black body with an increasing emission radius of the order of $10^{12}$ cm. Moreover, this component is
undetected in later XRT observations and it is interpreted as shock break out from a dense wind.
\subsection{\normalsize{XRF 050416A}}
\label{050416A}

\begin{figure}[ht]
   \centering
   \epsscale{0.7}
   \plotone{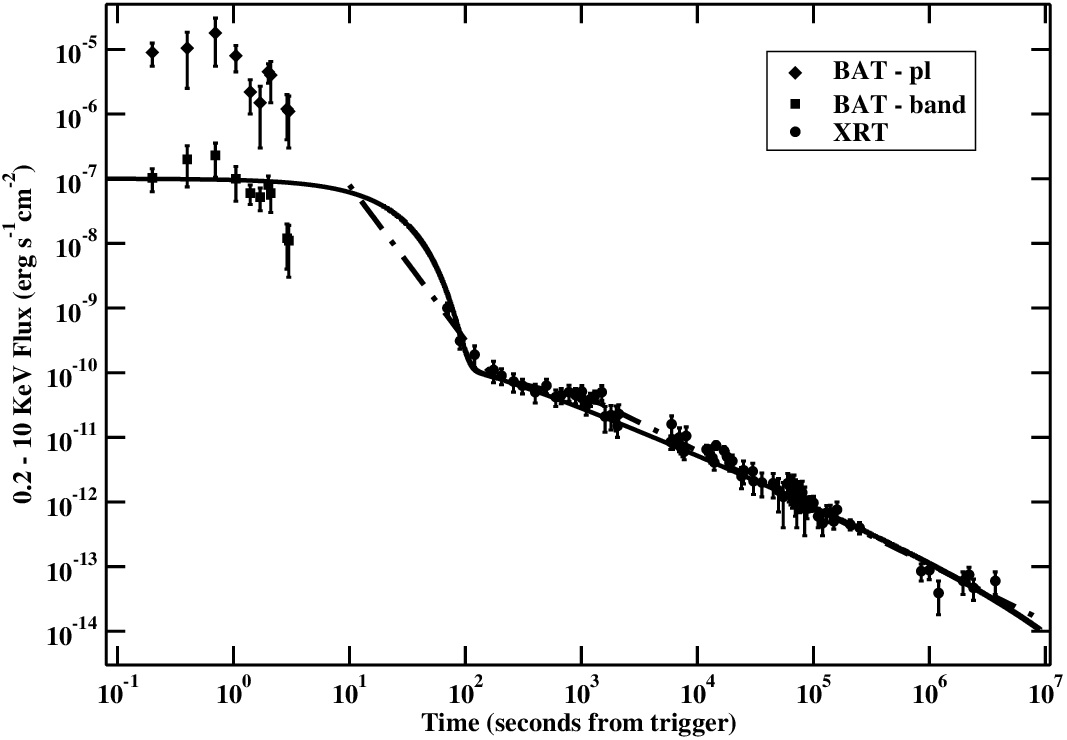}
\caption{\label{fig_12}
BAT and XRT  light curves  of XRF 050416A. The data was extracted from Fig.~2 in Mangano et al. (2006). The BAT light
curve was extrapolated into the XRT energy band with two different spectral models~\cite{Mangano:2006}. The dot-dashed
line is the phenomelogical broken power-law fit~\cite{Mangano:2006}. The full line is our light curve Eq.~(\ref{2.12})
with parameters in Eq.~(\ref{2.29}).}
\end{figure}

Swift discovered XRF 050416A with the Burst Alert Telescope on  2005 April 16. After about 76 seconds from the burst
trigger, XRT began collecting data~\cite{Mangano:2006}. The X-ray light curve was monitored up to 74 days after the onset
of the burst. The very soft spectrum of the burst classifies this event as an X-ray flash. \\
\indent
In Fig.~\ref{fig_12} we show the combined BAT-XRT light curve of XRF 050416A. The BAT light curve was extrapolated into
the $0.2 - 10 \, keV$ energy band assuming two different spectral law, the Band best fit model (full squares in
Fig.~\ref{fig_12}) and a single power law best fit model (full diamonds in Fig.~\ref{fig_12}). \\
\indent
From Fig.~\ref{fig_12} we see that the X-ray light curve initially decays very fast and subsequently flattens. It is
evident that the XRT light curve decay is not consistent with a single power law. Indeed, Mangano et al. (2006) found that
a doubly-broken power law improves considerably the fit of the light curve (dot-dashed line in Fig.~\ref{fig_12}). On the
other hand, we may adequately reproduce the combined BAT-XRT light curve with our light curve Eq.~(\ref{2.12}). To this
end, we assume that the early light curve is described by the BAT data extrapolated with the Band best fit model. By
fitting our Eq.~(\ref{2.12}) to the data, we find:
\begin{eqnarray}
\label{2.29}
 L_{G}(0) \; & \simeq & \; 1.0 \; 10^{-7} \; \frac{erg}{ cm^{2} \, sec} \; \; , \; \;
 \eta_{G} \; \simeq \;
             0.895 \; \; , \; \; t_{diss} \;  \simeq \;  1.8 \; 10^2 \; sec \;  \; \;  \\
\nonumber
 L_{S}(0) \; & \simeq & \; 2.5 \; 10^{-10}   \; \frac{erg}{ cm^{2} \, sec} \; ,  \;
  \eta_{S} \; \simeq \; 0.60 \; ,  \; \tau_{diss} \;  \simeq \; 8.0 \; 10^7 \; sec \;  ,
  \; \kappa  \; \simeq \; 2.0 \; 10^{-2} \; sec^{-1}  \; ,
\end{eqnarray}
Indeed, Fig.~\ref{fig_12} shows that our light curve is able to account for the light curve of XRF 050416A.
\section{\normalsize{ORIGIN OF GAMMA RAY BURSTS FROM P-STAR MODEL}}
\label{origin}
\begin{figure}[ht]
   \centering
   \epsscale{0.7}
   \plotone{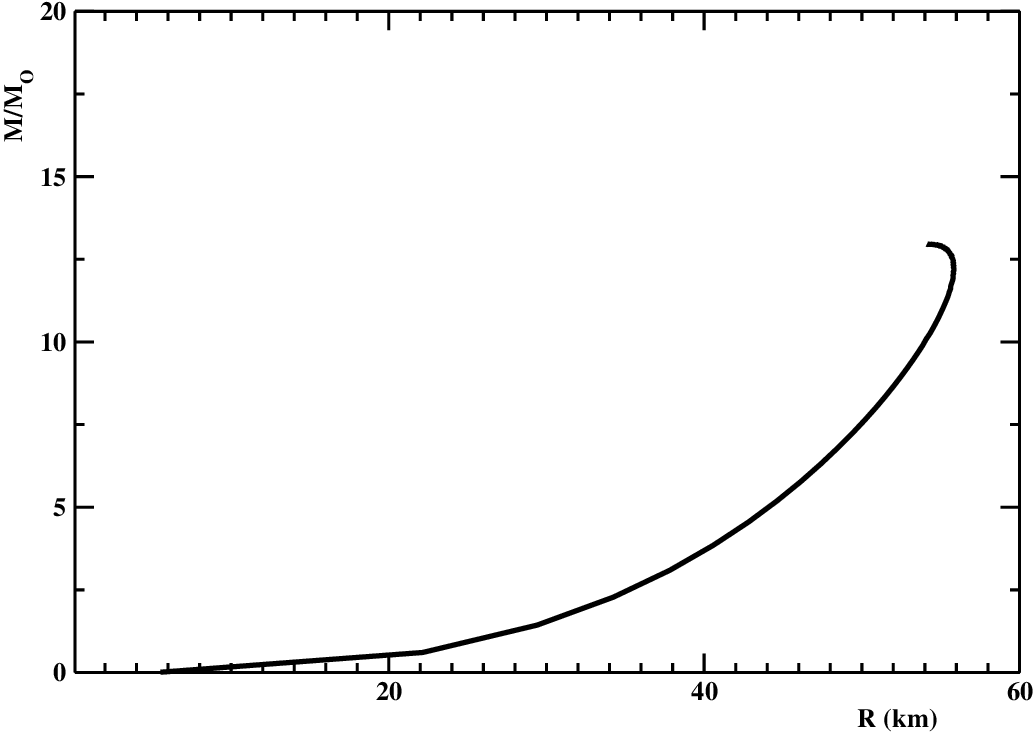}
\caption{\label{fig_13}
Gravitational mass M plotted versus stellar radius R for p-stars with $\sqrt{gH} = 0.1 \; GeV$.}
\end{figure}

The results in previous Section show that the light curves of Swift gamma ray bursts can be successfully described by the
approach developed in Cea (2006) to quantitatively account for light curves for both soft gamma repeaters and anomalous
X-ray pulsars. This leads us to suppose that the same mechanism is responsible for bursts from gamma ray bursts, soft
gamma repeaters, and anomalous X-ray pulsars. \\
\indent
In  Cea (2006) we showed that soft gamma repeaters and anomalous X-ray pulsars can be understood within our recent
proposal of p-stars, namely  compact quark stars in $\beta$-equilibrium with electrons in a chromomagnetic condensate (Cea
2004a,b). In particular, the bursts are powered by glitches, which in our model are triggered by dissipative effects in
the inner core. The energy released during a burst is given by the magnetic energy directly injected into the
magnetosphere:
\begin{equation}
\label{3.1}
\delta  E_{B}^{\emph{ext}} \; \simeq \; \frac{1}{3} \; R^3 \; B_S^2 \; \frac{\delta B_S}{ B_S} \; \; .
\end{equation}
For magnetic powered pulsars  with $M \, \sim \, M_{\bigodot}$ and $R \, \sim \, 10 \, Km$, we have $B_S \, \lesssim \,
10^{17} \, Gauss$. So that, from  Eq.~(\ref{3.1}) we get:
\begin{equation}
\label{3.2}
\delta  E_{B}^{\emph{ext}} \, \simeq \, 2.6 \, 10^{50} \, ergs  \, \left ( \frac{B_S}{10^{17} \, Gauss} \right )^2
                             \, \frac{\delta B_S}{ B_S} \, \lesssim \, 2.6 \, 10^{50} \, ergs \; \;  .
\end{equation}
The gamma-ray energy released in gamma ray bursts is narrowly clustered around $5.0 \, 10^{50} \, ergs$ \cite{Frail:2001}.
Thus, even though it is conceivable that a small fraction of gamma ray bursts could be explained by burst like the 2004
December 27 giant flare from SGR 1806-20, we see that canonical magnetic powered pulsars (canonical magnetars) do not
match the required energy budged to explain gamma ray bursts. On the other hand, we find that massive magnetars, namely
magnetic powered pulsars with $M \, \sim \, 10 \, M_{\bigodot}$ and $R \, \sim \, 10^2 \, Km$,
could furnish the energy needed to fire the gamma ray bursts. \\
\indent
The possibility to have massive pulsars stems from the fact that our p-stars do not admit the existence of an upper limit
to the mass of a completely degenerate configuration. In other words, our peculiar equation of state of degenerate up and
down quarks in a chromomagnetic condensate allows the existence of finite equilibrium states for stars of arbitrary mass.
In fact, in Fig.~\ref{fig_13} we display the gravitational mass versus the radius for p-stars with chromomagnetic
condensate $\sqrt{gH} = 0.1 \; GeV$.

Note that the strength of the chromomagnetic condensate of massive magnetars is reduced by less than one order of
magnitude with respect to canonical magnetars. Thus, we infer that for massive pulsars $B_S \, \lesssim \, 10^{16} \,
Gauss$. Using Eq.~(\ref{3.1}) we find:
\begin{equation}
\label{3.4}
\delta  E_{B}^{\emph{ext}} \, \simeq \, 2.6 \, 10^{51} \, ergs  \, \left ( \frac{B_S}{10^{16} \, Gauss} \right )^2
                             \, \frac{\delta B_S}{ B_S} \, \lesssim \, 2.6 \, 10^{51} \, ergs \; \;  ,
\end{equation}
that, indeed, confirms that massive magnetars are a viable mechanism for gamma ray bursts. \\
\indent
An interesting consequence of our proposal is that at the onset of the bursts there is an almost spherically symmetric
outflow from the pulsar, together with a collimated jet from the north magnetic pole~\cite{cea:2006}. Indeed, following
the 2004 27 December giant flare from SGR 1806-20 it has been detected a radio afterglow consistent with a spherical, non
relativistic expansion together with a sideways expansion of a jetted explosion. More interestingly, the lower limit of
the outflow energy turns out to be $E \gtrsim 10^{44.5} \; ergs$~\cite{Gelfand:2005}. This implies that the blast wave and
the jet may dissipate up to about $10 \, \% $ of the total burst energy. In the case of gamma ray bursts, according to our
proposal we see that at the onset of the burst there is a matter outflow with energies up to $\sim \, 10^{50} \, ergs$. We
believe that this could explain the association of some gamma ray bursts with supernova explosions.
\section{\normalsize{CONCLUSIONS}}
\label{conclusion}
Let us conclude by briefly summarizing the main results of the present paper. We have presented an effective theory which
allows us to account for X-ray light curves of both gamma ray bursts and X-ray rich flashes. We have shown that the
approach developed to describe the light curves from anomalous $X$-ray pulsars and soft gamma-ray repeaters works
successfully even for gamma ray bursts. This leads us the conclusion that the same mechanism is responsible for bursts
from gamma ray bursts, soft gamma repeaters, and anomalous X-ray pulsars. In fact, we propose that gamma ray bursts
originate by the burst activity from massive magnetic powered pulsars.
\end{document}